\documentclass[article,aps,pra,onecolumn,groupedaddress]{revtex4}
\usepackage{amsmath}    
\usepackage{graphicx}   
\usepackage{epsfig}

\begin{document}

\title{Decoy States and Two Way Quantum Key Distribution Schemes}
\author{J.S. Shaari$^{a}$, Iskandar Bahari$^{b}$ and Sellami Ali$^{b}$}
\address{\smallskip $^{a}${\textit{Faculty of Science, International Islamic University of Malaysia (IIUM), Jalan Bukit Istana, 25200 Kuantan, Pahang, Malaysia}} 
\\
$^{b}${\textit{Information Security Cluster, MIMOS Berhad, Technology Park Malaysia, 57000 Kuala Lumpur, Malaysia}}
}
\begin{abstract}
We study the possible application of the decoy state method on a basic two way quantum key distribution (QKD) scheme to extend its distance. Noting the obvious advantage of such a QKD scheme in allowing for single as well as double photon contributions, we derive relevant lower-bounds on the corresponding gains in a practical decoy state implementation using two intensities for decoy states. We work with two different approaches in this vein and compare these with an ideal infinite decoy state case as well as the simulation of the original.  
\end{abstract}

\maketitle
\section{Introduction}
\noindent Quantum mechanics which at one time challenged the beliefs of even scientists about nature has come today to be the corner stone of many if not all explanations of physical phenomenon at the atomic level. More than that, the theories have even seen applicability, mainly in the form of quantum computation \cite{deutsch, divincenzo, ekert} and quantum cryptography \cite{gisin}. The latter has seen the light of commercial realizations and to quite an extent has led itself into the world of engineering concerns where many challenges are at the level of implementation.  

Quantum cryptography, or more specifically, quantum key distribution (QKD) was arguably born in the seminal work of Bennet and Brassard \cite{bb84} with a protocol now popularly known as BB84. Twenty five years later today, the field has seen many inventions, improvements and of course plenty important physics discovered. However, despite the various protocols available today, its implementation is still unfortunately less than ultimately desirable. This is given the many imperfections in devices used in its realization, consequently limiting the effective distance to generate a secure key between two legitimate parties. Of many, it is arguably the case of the lack of a perfect single photon source technology. However, in \cite{hwang}, an interesting method called the decoy state method was proposed to compensate and since then has been studied somewhat extensively though limited mostly to BB84 \cite{bb84} and SARG04 \cite{SARG04} protocol \cite{X,sh,shc,sell,fred}. Before indulging in our intent in this paper, we will briefly mention another family of QKD protocol which can be described by its two way nature. Amongst the earlier works include \cite{bostroem,cai,deng,LM}. However, we will extensively make reference to the \cite{deng,LM} and refer to it as LM05 (this term was noted in \cite{LM05}).

The LM05 is analogous to the BB84 in the sense it uses 4 states derived from 2 mutually unbiased bases. It functions on the basis of sending a state prepared by one party, Bob to another, Alice and encoding is done by Alice using a unitary transformation (identity or a flip). The encoded qubit is then sent back to Bob and a sharp measurement may be made. To ensure security, a control mode, (completely analogous to BB84) where Alice would measure the received qubit, is executed randomly and results would be compared on a public channel. Given various results on specific attacks on the protocol, it is believed that the protocol has some security advantage over prepare and measure schemes. This was further noted in \cite{chiri} when considering the difficulty of cloning unitary transformations compared to states. In the case of implementation given imperfect devices, especially in consideration of the notorious imperfect single photon source, Lucamarini et al. \cite{marco} showed that the protocol exhibits a higher secure key rate given certain distances over the BB84; not surprisingly due to the fact that the LM05 inherently is more robust against a photon number splitting attack (a two photon pulse is not enough for an eavesdropper to perform such an attack). It is unfortunate however, that the two way channel for the protocol suffers losses on a quadratic scale  \cite{LM}.  

In this work, we will for the first time, we believe, consider the LM05 using the decoy method with the intention of extending the distance for a secure implementation. The flavor of improvement comes from the fact that LM05 may allow for key distillation even for 2 photons per-pulse as noted in \cite{marco} (which in this sense, similar to that of SARG04). In terms of a finite decoy state implementation, we propose two different approaches with the first (at least with regards to the double photon contribution) being an echo of the derivations in \cite{sh,shc,sell} for a two photon contribution. We note some issues relevant therein especially with regards to the lower bound for the gain of a double photon contribution, (referred to as $Q_2^L$)  which happens to be a function of the single photon contribution. In the second approach, in trying to avoid such dependencies of $Q_2^L$ on a single photon contribution term, we derive a different estimation for the lower bound for the relevant photon contributions and report the results for the different techniques. We then compare the distance for a secure key rate with that of a LM05 without decoy implementation. Finally we report on a brief comparison with a decoy state BB84 as well as the decoy state SARG04 protocol.

\section{Decoy State and LM05}

\noindent We will consider a simple fiber based setup with a weak pulse for a source. This allows for an immediate import of the model in \cite{X} with some differences, primarily coming from the two way nature of the protocol. If we consider the ideal case where an infinite number of decoy states are used, the conditional probability that Bob registers a detection given a pulse with $i$ photons was sent (usually referred to as the yield of the \textit{i-th} photon) as well the relevant quantum bit error rate (QBER) may be estimated respectively  by \cite{X} 
\begin{eqnarray}
Y_i=Y_0+\eta_i-Y_0\eta_i
\label{perfect}
\end{eqnarray}
\begin{eqnarray}
e_i=\dfrac{e_0Y_0+e_{detector}\eta_i}{Y_i}.
\end{eqnarray}
where the transmittance, $\eta_i=1-(1-\eta)^i$, $\eta=t\eta_{AB}$, $t=10^{-\alpha 2l/10}$. $\eta_{AB}$ is the efficiency related to measurement as well as encoding apparatus (for simplicity we assume the encoding process is done with unit efficiency) and we note the transmittance $t$ acquires a $2l$ term, i.e. double length $l$ term due to the two way nature of the protocol concerned \cite{marco}. The term $\alpha$ is the loss coefficient which reflects the losses in the channel and is measured in dB/km. The gain for the \textit{i-th} and the total gain with a source of intensity $\mu$ are given by $Y_i\exp{\left(-\mu\right)}\mu^i/i!$ and $Q_{\mu}=Y_0+1-\exp{\left(-\eta\mu\right)}$ \cite{X}.  
However, for a practical setup where finiteness is a reality, we consider the use of 2 decoy states with intensities $\nu_1$ and $\nu_2$ respectively and a signal state with intensity $\mu$ and reiterate the conditions in \cite{X}, $\nu_1+\nu_2<\mu$ and $\nu_1+\nu_2<1$. However, these conditions do not imply any constraint on the value of $\mu$, which must be ascertained only through maximizing the key rate. We bear in mind, the gains for the 2 decoy states and signal state, $Q_{\nu_1},Q_{\nu_2}$ and $Q_\mu$ respectively, may be observed experimentally.

The single photon contribution may be imported directly from \cite{X}. In calculating the double photon detection probability, it is tempting to adopt immediately the work of \cite{sh,shc,sell} for SARG04. Before we proceed, let us note the formula derived for double photon detection in \cite{sh,shc} contains a term for the lower bound of a single photon contribution of the form $-Y_1 \eta_A f(\nu_1,\nu_2,\mu)$. As the function $f(\nu_1,\nu_2,\mu)>0$ in \cite{shc} (in \cite{sh}, this can also achieved, especially when $\mu>\nu_1+\nu_2$), in our opinion, this would put some questions on the `lower bound' nature of the double photon formula if $Y_1$ is not taken as the upper bound for the single photon contribution (this is especially given that $Y_1$ itself is not directly measured in experiments and only its lower bound may be estimated). In this section, we shall derive a similar formula with further scrutiny on the lower bound of $Y_2$. We begin by borrowing from \cite{X,sh,shc,sell}, we write
\begin{equation}
Q_{\nu_1} e^{\nu_1}-Q_{\nu_2} e^{\nu_2}=\sum_{n=0}^2 Y_i \dfrac{(\nu_1^i-\nu_2^i)}{i!}+\sum_{n=3}^\infty Y_i \dfrac{(\nu_1^i-\nu_2^i)}{i!}
\end{equation}
The second term in the right hand side of the equation highlights our interest in single as well as double photon contribution. 
Writing 
\begin{eqnarray}
\sum_{n=3}^\infty Y_i \dfrac{\mu^i}{i!}=Q_\mu e^{\mu}-\sum_{n=0}^2 Y_i \dfrac{\mu^i}{i!}
\end{eqnarray} we may thus note
\begin{eqnarray}
Q_{\nu_1} e^{\nu_1}-Q_{\nu_2} e^{\nu_2}&\leq&\sum_{n=0}^2  \dfrac{Y_i(\nu_1^i-\nu_2^i)}{i!}+\dfrac{\nu_1^3-\nu_2^3}{\mu^3}\sum_{n=3}^\infty Y_i \dfrac{\mu^i}{i!}\\\nonumber
&=&\sum_{n=0}^2 Y_i \dfrac{(\nu_1^i-\nu_2^i)}{i!}+\dfrac{\nu_1^3-\nu_2^3}{\mu^3}\left(Q_\mu e^\mu-Y_0-\sum_{n=1}^2 Y_i \dfrac{\mu^i}{i!}\right)
\label{root}
\end{eqnarray} 
Such an inequality is derived from the fact that $a^3-b^3>a^i-b^i, \forall i>3$ (we omit the proof for this). 
Thus
\begin{eqnarray}
Y_2\geq Y_2^L=\dfrac{2\mu \left\{Q_{\nu_1} e^{\nu_1}-Q_{\nu_2} e^{\nu_2}-\left[Y_1 \left(\dfrac{\mu^2(\nu_1-\nu_2)}{\mu^2}-\dfrac{\nu_1^3-\nu_2^3}{\mu^2}\right)+\dfrac{\nu_1^3-\nu_2^3}{\mu^3}\left(Q_\mu e^\mu-Y_0^L\right)\right]\right\}}{(\nu_1^2-\nu_2^2)\mu-\nu_1^3+\nu_2^3}  
\end{eqnarray}
where $Y_0^L$ is given by \cite{X}
\begin{eqnarray}
Y_0^L=\max\left\{\dfrac{\nu_1Q_{\nu_2}e^{\nu_2}-\nu_2Q_{\nu_1}e^{\nu_1}}{\nu_1-\nu_2},0\right\}%
\end{eqnarray}
If we consider cases where $\mu>\nu_1+\nu_2$, then the numerator for the coefficient of $Y_1$,
\begin{eqnarray}
\mu^2(\nu_1-\nu_2)-\left(\nu_1^3-\nu_2^3\right)&>&(\nu_1+\nu_2)^2(\nu_1-\nu_2)-\left(\nu_1^3-\nu_2^3\right)\\\nonumber
&=&\nu_1\nu_2(\nu_1-\nu_2)>0
\end{eqnarray}
As there is no way of ascertaining the exact value of $Y_1$ (one only ascertains its lower bound \cite{X}), mathematically speaking, the term $Y_1$ cannot be taken as anything other than an upper bound lest $Y_2^L$ effectively does not become a lower-bound. Such a concern is absent in considering an lower bound for $Y_1$ in \cite{X} as it can be expressed with no explicit dependency on terms other than $Y_0$. It is tempting to immediately use the value $Y_1$ derived from eq.(\ref{perfect}) (this was the choice that was used in \cite{sh,shc}), which would be valid in the absence of any Eve's tempering. Let us for brevity call this $Y_1^{\infty}$ (the superscript $\infty$ is to reflect infinite decoy states used in the estimation). The nagging question is the following; can $Y_1>Y_1^{\infty}$? We may sweep this under the rug by ignoring the possibility. However for the purpose of being rigorous, we need to understand what values $Y_1$ may achieve beyond $Y_1^{\infty}$. This may be seen as Eve's tempering.  We consider the following inequality immediately implied from eq.(3)
\begin{eqnarray}
Q_{\nu_1} e^{\nu_1}-Q_{\nu_2} e^{\nu_2}\geq\ Y_1\left(\nu_1-\nu_2\right)+\dfrac{Y_2\left(\nu_1^2-\nu_2^2\right)}{2}
\label{eq9}
\end{eqnarray}
where the values of $Y_1$ and $Y_2$ are to some extent subject to Eve's influence and an equality only holds when pulses containing more than 2 photons are blocked. 
The dependency on $Y_2$ tells us that one may even go as far as considering $Y_2=0$ zero to ensure that $Y_1$ is `genuinely' upper-bounded. It is quite evident to see that some trivial algebraic manipulation may provide with a channel loss coefficient $\alpha>0$ for such an equality to hold (our reference to $\alpha$ is made in view of Eve's `accessible loss' \cite{lutj}).  While this is certainly more reasonable from a strict rigorous perspective (to respect the requirement of an upper-bound for $Y_1$), it lacks physical appeal as it requires Eve's tempering to include an increasing of the single photon gain, which seems counter-intuitive in the face of any conservative attack by Eve. The relevant formulas are then
\begin{eqnarray}
Y_1^L=\dfrac{\mu}{\mu\left(\nu_1-\nu_2\right)-\nu_1^2+\nu_2^2}\left[Q_{\nu_1} e^{\nu_1}-Q_{\nu_2} e^{\nu_2}-\dfrac{\nu_1^2-\nu_2^2}{\mu^2}\left(Q_\mu e^\mu- Y_0^L\right)\right]%
\end{eqnarray}
\begin{eqnarray}
Y_2^L=\dfrac{2\mu\left\{Q_{\nu_1} e^{\nu_1}-Q_{\nu_2} e^{\nu_2}-\left[Y_1^{U} \left(\dfrac{\mu^2(\nu_1-\nu_2)}{\mu^2}-\dfrac{\nu_1^3-\nu_2^3}{\mu^2}\right)+\dfrac{\nu_1^3-\nu_2^3}{\mu^3}\left(Q_\mu e^\mu-Y_0^L\right)\right]\right\}}{(\nu_1^2-\nu_2^2)\mu-\nu_1^3+\nu_2^3}  
\end{eqnarray}
where $Y_1^L$ is adopted from \cite{X}. These can then be applied to find $Q_1^L$ and $Q_2^L$. The upper bound for the single photon yield, $Y_1^U$ can be taken as $Y_1^{\infty}$ or for a more genuine upper bound, the case for $Y_2^L=0$.  
We believe the construction of the inequalities above to be mathematically sound though an equalities would physically reflect changes in the photon statistics. Let us note that similar arguments were made for $Y_2$ in \cite{X} when deriving $Y_1^L$.   

In deriving the upper bounds for the errors $e_1$ and $e_2$ relevant to $Y_1$ and $Y_2$ respectively, we write out the equations for errors assuming they come only from single, double photon contribution as well as dark pulse. We do note though our derivation conveniently forces the $Y_0$ term to vanish. 
\begin{eqnarray}
E_{\nu_1}Q_{\nu_1} e^{\nu_1}=e_0Y_0+e_1Y_1\nu_1+e_2Y_2\dfrac{\nu_1^2}{2}\\\nonumber
E_{\nu_2}Q_{\nu_2} e^{\nu_2}=e_0Y_0+e_1Y_1\nu_2+e_2Y_2\dfrac{\nu_2^2}{2}\\\nonumber
E_{\mu}Q_{\mu} e^{\mu}=e_0Y_0+e_1Y_1\mu+e_2Y_2\dfrac{\mu^2}{2}
\end{eqnarray} 
which we work out to be
\begin{eqnarray}
E_{\nu_1}Q_{\nu_1} e^{\nu_1}-E_{\nu_2}Q_{\nu_2} e^{\nu_2}=e_1Y_1\left(\nu_1-\nu_2\right)+e_2Y_2\dfrac{\nu_1^2-\nu_2^2}{2}
\label{13}
\end{eqnarray}
\begin{eqnarray}
E_{\mu}Q_{\mu} e^{\mu}-E_{\nu_2}Q_{\nu_2} e^{\nu_2}=e_1Y_1\left(\mu-\nu_2\right)+e_2Y_2\dfrac{\mu^2-\nu_2^2}{2}
\label{14}
\end{eqnarray} 
We then solve for $e_1$ by eq.(\ref{13})$\times \left(\mu^2-\nu_2^2\right)-$eq.(\ref{14})$\times \left(\nu_1^2-\nu_2^2\right)$ and $e_2$ by eq.(\ref{13})$\times \left(\mu-\nu_2\right)-$eq.(\ref{14})$\times \left(\nu_1-\nu_2\right)$. This gives the following upper-bounds for the respective errors
\begin{eqnarray}
e_1^U=\dfrac{\left(E_{\nu_1}Q_{\nu_1} e^{\nu_1}-E_{\nu_2}Q_{\nu_2} e^{\nu_2}\right)\left(\mu^2-\nu_2^2\right)-\left(E_{\mu}Q_{\mu} e^{\mu}-E_{\nu_2}Q_{\nu_2} e^{\nu_2}\right)\left(\nu_1^2-\nu_2^2\right)}{Y_1^L\left[\left(\nu_1-\nu_2\right)\left(\mu^2-\nu_2^2\right)-\left(\mu-\nu_2\right)\left(\nu_1^2-\nu_2^2\right)\right]}
\end{eqnarray}
\begin{eqnarray}
e_2^U=\dfrac{\left(E_{\nu_1}Q_{\nu_1} e^{\nu_1}-E_{\nu_2}Q_{\nu_2} e^{\nu_2}\right)\left(\mu-\nu_2\right)-\left(E_{\mu}Q_{\mu} e^{\mu}-E_{\nu_2}Q_{\nu_2} e^{\nu_2}\right)\left(\nu_1-\nu_2\right)}{Y_2^L\left[\dfrac{\left(\nu_1^2-\nu_2^2\right)}{2}\left(\mu-\nu_2\right)-\dfrac{\left(\mu^2-\nu_2^2\right)}{2}\left(\nu_1-\nu_2\right)\right]}
\end{eqnarray}

Before we proceed with relevant key rate curves for the above bounds, we will in the following section, consider a slightly different approach to derive a lower bound for $Y_1$ and $Y_2$ at the same time.    

\section{Lower-bound of $Y_1+Y_2$}
While deriving a lower bound for $Y_1$ is somewhat dependent only on manipulations of multiphoton pulses, that of $Y_2$ is a function of $Y_1$ as well (as described in the previous section). In this section we will consider a variation in deriving the lower bound for the key rate for LM05. The method will be based on trying to achieve the lower bound of $Y_1$ and $Y_2$ at the same time as opposed to the above. In other words, we should hope to derive a lower bound for $Y_1+Y_2$. Let us consider the following 
\begin{eqnarray}
Y_1\mu+\dfrac{Y_2}{2}\mu^2=\dfrac{\left(Y_1+Y_2\right)}{2}\mu^2+\left(Y_1\mu-\dfrac{Y_1\mu^2}{2}\right)
\end{eqnarray}
and
\begin{displaymath}
\sum_{i=1}^2 Y_i \dfrac{(\nu_1^i-\nu_2^i)}{i!}< \sum_{i=1}^2 Y_i (\nu_1-\nu_2)
\end{displaymath}
thus from eq.(\ref{root}) 
\begin{eqnarray}
Q_{\nu_1} e^{\nu_1}-Q_{\nu_2} e^{\nu_2}<\sum_{i=1}^2 Y_i (\nu_1-\nu_2)+\dfrac{\nu_1^3-\nu_2^3}{\mu^3}\left(Q_\mu e^\mu-Y_0-\dfrac{\left(Y_1+Y_2\right)}{2}\mu^2-\left(Y_1\mu-\dfrac{Y_1\mu^2}{2}\right)\right). 
\end{eqnarray}  
Thus a lower bound for $Y_1+Y_2$ is given as
\begin{equation}
(Y_1+Y_2)^L=\dfrac{\left[Q_{\nu_1} e^{\nu_1}-Q_{\nu_2} e^{\nu_2}-\dfrac{\nu_1^3-
\nu_2^3}{\mu^3}\left(Q_\mu e^\mu- Y_0^L-\left(Y_1^L\mu-\dfrac{Y_1^L\mu^2}{2}\right)\right)\right]}{\left(\nu_1-\nu_2-\dfrac{\nu_1^3-\nu_2^3}{2\mu}\right)}
\end{equation}
Writing the contribution from the single and double photon gain, $\mathcal{Q}_{12}\left(\mu\right)$ may be given by 
\begin{eqnarray}
\mathcal{Q}_{12}\left(\mu\right)&=&Y_1e^{-\mu}\mu+\dfrac{Y_2}{2}e^{-\mu}\mu^2\\\nonumber
&=&\dfrac{\left(Y_1+Y_2\right)}{2}e^{-\mu}\mu^2+\left(Y_1\mu-\dfrac{Y_1\mu^2}{2}\right)e^{-\mu}
\end{eqnarray} 
we thus define the lower bound on the effective gain as 
\begin{equation}
\mathcal{Q}_{12}^L\left(\mu\right)=\left[\dfrac{\left(Y_1+Y_2\right)^L}{2}\mu^2+\left(Y_1^L\mu-\dfrac{Y_1^L\mu^2}{2}\right)\right]e^{-\mu}
\end{equation} 
In this way, rather than consider the rates $Y_1$, then $Y_2$ separately (correspondingly the gains) we lump them as an effective gain. We do admit that there is the need for $Y_1^L$, though as it is accompanied with a positive coefficient, it does not carry the `upper bound' issue of the earlier section. In principle, one could have exclusively estimated for $Y_1+Y_2$, though our earlier estimation in that vein (not presented) had provided much poorer key rates. We observed that a derivation making use of the lower bound for $Y_1$ of eq.(10) improved the key rate estimation. 

Next, we consider the errors in the channel. If we assume all errors come from single and double photon detection (as well as dark counts), we could consider 
\begin{eqnarray}
E_\mu Q_\mu =e_0 Y_0e^{-\mu}+\mathcal{E}\mathcal{Q}_{12}\left(\mu\right)
\end{eqnarray}
where $\mathcal{E}$ is the effective error for the effective critical gain. Thus we may upper bound the error, $\mathcal{E}^U$
\begin{eqnarray}
\mathcal{E}^U=\dfrac{E_\mu Q_\mu-e_0 Y_0^Le^{-\mu}}{\mathcal{Q}_{12}^L}
\end{eqnarray}
This derivation has the advantage that one does not concern oneself with exactly how Eve may manipulate the single photon yield and its influence on the double photon contributions individually as long as $\mathcal{Q}_{12}^L$ is respected. 
As the method here addresses the inclusion of a double photon yield, one may consider the possibility of a similar approach to be done for SARG04. However, as the error rates from single and double photon contributions require different amount of bits to be discarded in privacy amplification, one must, upon ascertaining the effective error rate (which combines that from single as well as double photon contribution) determine the amount of bits to be discarded by assuming that it is either fully resulting from single or double photon contribution. We refer to \cite{fred} for the relevant formulas. The number of bits based on the larger of the two should then be discarded to assure no underestimation of Eve's information had been made. 

\subsection{Key Rate Formulas}
This is not really as straightforward as one would hope for when it comes to two way QKD protocols. The reasoning is that, while much work for a complete security proof has been done for prepare and measure schemes (BB84 as the pioneering example), literally none has been for the case of LM05. As a matter of fact, to date, to the best of our knowledge, the security of such protocols are studied in the context of selected types of attacks and thus lacks a proper sense of generality. The only two way QKD scheme receiving a very extensive treatment in terms security analysis would be in \cite{piran}. This is however in the context of continuous variables addressing all types of collective Gaussian attack. 

Nevertheless, to consider a fair comparison with the work in \cite{marco} we will use the security as defined in the context of individual attacks; an adoption of Lutkenhaus's work \cite{lutk}. It is instructive to note that while Lutkenhaus's work addressed single photon contribution to a secure key, the framework may immediately be extended to include a double photon contribution provided the amount of bits discarded for privacy amplification in the double photon case do not exceed that for the single photon case. We will provide a heuristic justification for this in the following paragraph. Thus, extending from BB84 version in \cite{phd} and SARG04's \cite{sh,shc,sell}, we will consider the key rate formula
\begin{eqnarray}
R_{\infty}\geq -Q_\mu f(E_\mu)H(E_\mu)+\sum_{i=1}^2 Q_i[1-\tau(e_i)]
\label{r1}
\end{eqnarray}
and in the practical case, for the above two different approaches respectively
\begin{eqnarray}
R\geq -Q_\mu f(E_\mu)H(E_\mu)+\sum_{i=1}^2 Q_i^L[1-\tau(e_i)]
\label{r2}
\end{eqnarray}
\begin{eqnarray}
R_{12}\geq -Q_\mu f(E_\mu)H(E_\mu)+\mathcal{Q}_{12}^L[1-\tau(\mathcal{E}^U)]
\label{r3}
\end{eqnarray}  
where the amount of bits lost due to privacy amplification is given by $\tau(e)=\log_2{\left(1+4e-4e^2\right)}$ for $e<1/2$ and $1$ otherwise. $f(E_\mu)$ reflects the efficiency of the error correcting procedure (for simplicity we assume it to be 1.22, thus providing for a more pessimistic scenario). $H(e)$ is the binary entropic function given by $-e\log_2{(e)}-(1-e)\log_2{(1-e)}$. It is instructive to note despite the  necessary similarity and reference of the above formulas to \cite{sh,shc,sell}, the formula for the number of bits discarded for privacy amplification in \cite{sh,shc} are as prescribed in \cite{fred} (which were derived exclusively for SARG04). The work in \cite{sell} considers a simpler approach, where bits to be discarded are given as binary entropic functions of the bit errors noted in single and double photon pulses respectively. It is not clear if these formulas can be used for LM05 (probably not though). As we are considering an individual attack framework (not unlike \cite{lutk}), our choice of function reflecting bits discarded in privacy amplification is $\tau(e)$. Before further justifications, let us recall from \cite{lut99} that the function $\tau$ was derived from the optimization of the collision probability on the reconciled key given Eve's general but individual attack. While it was originally considered only for Eve's attack on a single photon pulse, we will argue that one may use it even for a double photon pulse under certain assumptions. The following argument, which is constructed in a hypothetical framework is really aimed at upper-bounding Eve's information of a key for both single and double photon pulses.  

Heuristically one may imagine a hypothetical scenario where Eve is provided with some `privileged' information depending on the number of photons in a pulse sent after all communications between Alice and Bob  have been completed; the analysis may then be simplified. The only constraint we prescribe for this privileged information is that it should be made available to Eve \textit{only after} the whole communication between Alice and Bob has ended (at least the quantum phase). Say if a single photon state was sent, Eve is later privileged to the knowledge of the state originally sent by Bob and in the event of a double photon state, only the basis of the state is divulged. The protocol then reduces to a BB84 like setup, where Bob estimates the state Alice sends him by making a measurement in the correct basis in the face of disturbance. Eve thus needs to attack only the photon traversing from Alice to Bob after the encoding in the backward path. If it was a double photon state, she would have hijacked one of the photons in the pulse in the forward path and at the end, may choose the relevant basis to measure while causing minimal disturbance to the system. As she knows the state of the qubit in the forward path (or the basis to measure a hijacked qubit), her knowledge of Alice's encoding would simply depend on her estimation of the state she attacked, which is only of one photon. This can be seen more clearly as follows: Eve's probability of guessing the state in the forward path becomes unity whilst in the backward path is dependent on her attack strategy, say $\mathcal{F}$. Her probability of guessing Alice's encoding correctly in both types of pulses thus becomes $1\cdot\mathcal{F}$, effectively identical to a BB84 individual attack. From an information theoretic perspective, her Shanon information gain of Alice's encoding should be given by $1-H\left(\mathcal{F}\right)$, where $H$ is the binary entropic function. With regards to privacy amplification, as $\tau$ in \cite{lutk} is derived from the collision probability on a reconciled key between Alice and Bob, which is essentially a function of $\mathcal{F}$, in either of the two cases of our concern, Eve effectively attacks only one photon and the amount of bits to be discarded by the legitimate parties should therefore be given by $\tau$. The argument here is certainly less realistic as it is quite unimaginable why such information is leaked to Eve; nevertheless, as it involves much privileged information provided to Eve, presumably in her favor, her actual information of the key should probably be less than depicted above. Hence we believe that the key rate formula used is a pessimistic estimation for LM05.
\section{Results and discussions}

\noindent Given formulas (\ref{r1})-(\ref{r3}), we may plot the relevant curves for secure key rate against distance. For the purpose of definiteness in calculations, we chose parameters derived from the GYS experiment \cite{gys}. Let us first describe the curves plotted in figure 1; a curve for infinite decoy state, $R_{\infty}$ is plotted using eq.(\ref{r1}); two curves, $R_{1+2}$, $R_{U}$ uses eq.(\ref{r2}) with  differing considerations on $Y_1^U$. The first of the two applies $Y_1^U=Y_1^\infty$ and the second considers the `genuine' upper-bound of $Y_1$ where $Y_2^L=0$ respectively. A fourth curve, $R_{12}$, is plotted using eq.(\ref{r3}) and the final curve, $R_{LM05}$ for a non-decoy case, we consider the formula given in \cite{marco}.

\begin{figure}[htbp]
	\centering
			\includegraphics[angle=0,width=0.80\textwidth]{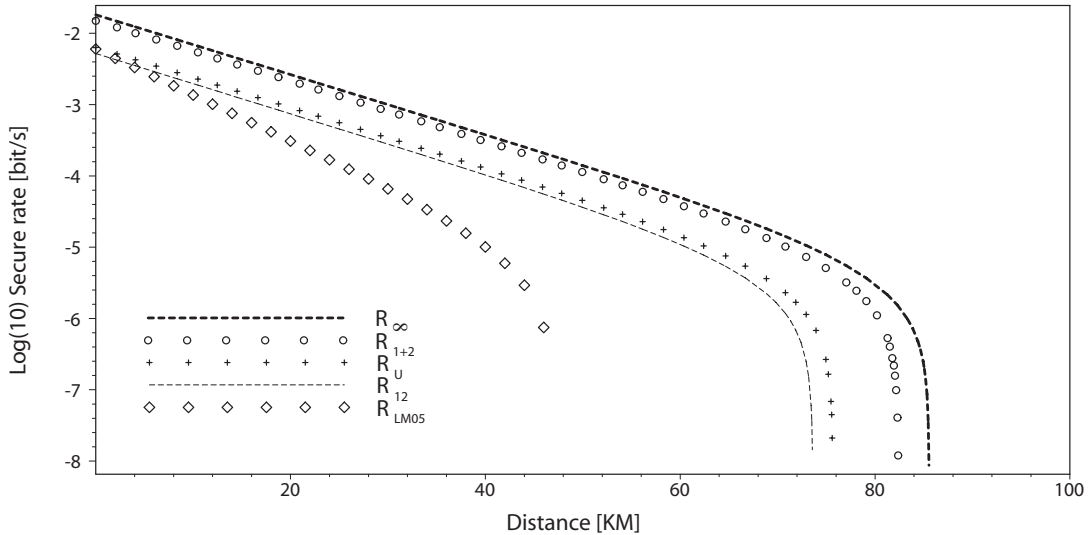}
		\caption{The above shows five key rate curves, $R_{\infty}$,$R_{1+2}$,$R_{U}$ and $R_{12}$ with the description in the text. In our plots for finite decoy states, we had chosen the values of $\nu_1=0.05, \nu_2=0$. As for the signal intensity, $\mu=0.45$ was used for both $R_{\infty}$ and $R_{1+2}$. $\mu=0.45$ and $0.35$ were used for $R_U$ and $R_{12}$ respectively. }
\end{figure}

Let us begin by comparing the lower bound for the key rate calculated using the different methods above.
From figure 1, we note that the key rates $R_{\infty}$ and $R_{1+2}$ ($Y_1^\infty$ are used as upper bound for $Y_1$) tends to be essentially the same for most of the distances. A more pessimistic approach is seen in $R_U$ ($Y_2^L=0$) which is equivalent to the case where the double photon contribution is excluded. Let us note at passing, we do not think this is the scenario where one should be much concerned with as argued earlier though it suggests the significance of the two photon contribution. This may be further seen from our choice of $\mu=1.16$ for the key rates of infinite decoy case and the case where $Y_1^U=Y_1^\infty$. The choice of a $\mu>1$ is hardly surprising if we consider $\sum_{i=1}^2(\exp{(-\mu)})\mu^i/i!$, i.e. the sum of probabilities for a single and double photon contribution in a weak source. Solving for the first derivative equaling zero, $e^{-\mu}(2-\mu^2)/2=0$, we readily get a $\mu=\sqrt{2}$ as the maximum point which is greater than the maximum $1$ for a single photon contribution only. However the lowest estimation for a secure key is noted for our second approach, $R_{12}$, which admittedly is a little disappointing. A more detailed study regarding the behaviors of the key rate estimation formulas under limiting cases for the decoy state intensities should help illuminate the differences between them, i.e. which would present a better approximation to a tight bound with regards to the infinite decoy case. Nevertheless, given the results at the moment, we are prone to think that a more refined method though based on the first should be considered as a proper estimation for a finite decoy implementation. On the whole, our practical approach does expectedly fall a little short with regards to the ideal infinite decoy case, where the distance is about 90km or so.

Comparison with the case where one does not employ the decoy state method, $R_{LM05}$ (our plot for the non-decoy case is an optimized version where numerics were used to determine the value of $\mu$ to achieve the largest key rate at intervals of distance) clearly illustrates how the use of decoy state increases the possible distance for a secure key rate almost by a factor of two. This is somewhat encouraging. 

In comparing LM05 to two protocols namely BB84 and SARG04, we consider the case for an infinite decoy scenario. In the former, we employ the key rate formula from \cite{X} and for the latter as described in \cite{fred}. The comparison has LM05 displaying a more favorable key rate for distances up to almost 36 km against BB84, while for SARG04, its advantage is displayed up to a distance of about 80 km. Understandably, it suffers losses greatly as compared to both BB84 and SARG04 after that. This can be seen in figure 2. We note that the plot uses (numerically) optimized intensity at each distance.      
\begin{figure}[htbp]
	\centering
			\includegraphics[angle=270,width=0.80\textwidth]{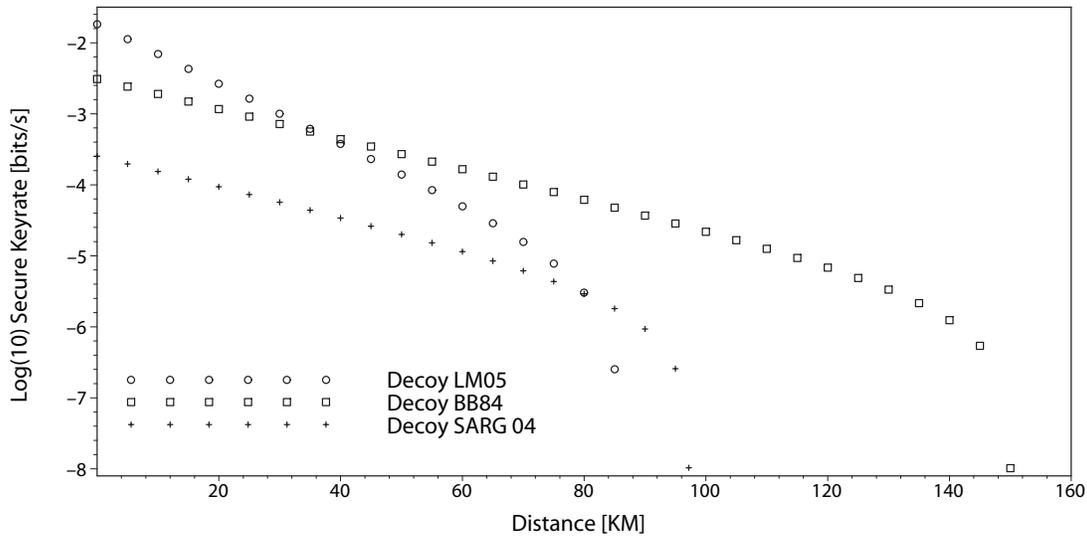}
	\label{fig:Fig3e}
	\caption{The above shows the comparison for the decoy based LM05, BB84 and SARG04 with decoy implementation. All three key rates are calculated using the infinite key rate formula.}
\end{figure}
We should point out that the comparisons with BB84 and SARG04 does not reflect conclusiveness. For one, the formulas for the latter two are based on unconditional security proofs and therefore promises unconditional secure keys while for LM05, the formula is based on independent attacks. Nevertheless given the fact that little can be gained by considering the independent attack for BB84 \cite{phd}, as well as a pessimistic case considered for LM05, we believe the sketch for the comparison is reasonable.   
Thus, we remain confident that what is presented is very much the pessimistic case for the LM05 due to the relatively simple and almost naive treatment for a secure key rate formula. It is known that in two way deterministic protocols like LM05, there is an asymmetry between the information shared between Alice and Eve, Bob and Eve and Alice and Bob. The deliberations above simply do not take this into account, i.e., as noted in \cite{marco}, in the case for LM05, one could have considered a reverse error correction which may give two way deterministic protocols a better gain.    

\section{Conclusion}
\noindent While double photon contributions are not deemed to be very significant for a secure key rate in SARG04 \cite{fred}, we believe that the case differs for a two way QKD protocol like LM05. This is given the fact that the errors from a single and two photon contributions are treated in an identical manner. We see that the finite decoy state case considered is quite encouraging and should justify further study on the matter. This should include a proper optimization with regards to the intensities as well as an optimal number of decoy states to be used. On a more fundamental level, it is noteworthy that LM05, has at its core the issue of how encoding is really `in' the unitary transformations rather than the states sent, hence Eve's estimations of states traveling to and fro in the channel is not immediately equal to the encoding. Rather, she only gets a correct estimation of the encoding with a probability of $F^2+(1-F)^2$ \cite{LM}, where $F$ is the probability of estimating a state correctly (this is assuming an attack where Eve measures, even with assistance of ancilla, the qubit sent from Bob to Alice independently of it as it traverses from Alice back to Bob). Given the domain of $F\in [0.5,1]$, we note her estimations as a convex function, thus suggesting a higher tolerance towards errors when compared to BB84.  It is also not clear if delayed measurements would even benefit Eve given the absence of bases revelation (relevant to signal states/ encoding and decoding of qubits) in LM05. A more detailed enterprise should see a possibly different formula to be applied in place of $\tau$ for bits to be discarded due to privacy amplification. Hence the study is still very much subject to improvement.

As far as physical realizations are concerned, given the already available experiments on LM05 \cite{marco}, an implementation of the decoy state is certainly feasible.
\newline
\newline
\noindent We would like to thank Marco Lucamarini, Faroukh Mukhamedov, Ressa Said and Chi-Hang Fred Fung for helpful discussions. We would also like to thank the reviewers for their most helpful comments allowing us to improve the work.


\section{References}
\begin{thebibliography}{99}
\bibitem{deutsch} D. Deutsch, Proceedings of the Royal Society of London A 400, (1985).
\bibitem{divincenzo}  D P DiVincenzo, Science 270, 5234, (1995).
\bibitem{ekert} A. Ekert and R. Jozsa, Rev. Mod. Phys. 68, 733 (1996).
\bibitem{gisin} N. Gisin, G. Ribordy, W. Tittel, and H. Zbinden, Rev. of Mod. Phys. 74, 145 (2002).
\bibitem{bb84} C. H. Bennett and G. Brassard, in Proc. of IEEE Int. Conference on Computers, Systems,
and Signal Processing (Bangalore, India, 1984), pp. 175 - 179.
\bibitem{hwang} W.-Y. Hwang, Phys. Rev. Lett. 91, 057901 (2003)
\bibitem{SARG04} V. Scarani, A. Acin, G. Ribordy, and N. Gisin, Phys. Rev. Lett. 92, 057901 (2004).
\bibitem{X} Xiongfeng Ma, Bing Qi, Yi Zhao, and Hoi-Kwong Lo,arXiv:quant-ph/0503005v5 10 May 2005
\bibitem{sh} ShengLi Zhang, XuBo Zou, ChenHui Jin and GuangCan Guo, arXiv:0807.1760v1 [quant-ph] 11 Jul 2008
\bibitem{shc} ShengLi Zhang, XuBo Zou , ChuanFeng Li, ChenHui Jin and GuangCan Guo, Chinese Science Bulletin 54, 11 (2009)
\bibitem{sell} Jing-Bo Li and Xi-Ming Fang, arXiv:quant-ph/0509077 v2 23 Sep 2005
\bibitem{fred} Chi-Hang Fred Fung, Kiyoshi Tamaki and Hoi-Kwong Lo, Phys. Rev. A 73 012337 (2006).
\bibitem{bostroem} K. Bostroem, T. Felbinger, Phys. Rev. Lett. 89 (2002) 187902.
\bibitem{cai} Q.-Y. Cai, B.W. Li, Chin. Phys. Lett. 21 (2004) 601.
\bibitem{deng} F.-G. Deng, G.L. Long, Phys. Rev. A 70 (2004) 012311.
\bibitem{LM} M. Lucamarini, S. Mancini, Phys. Rev. Lett. 94 (2005) 140501.
\bibitem{LM05} Alessandro Cere, Marco Lucamarini, Giovanni Di Giuseppe, and Paolo Tombesi, Phys. Rev. Lett. 96, 200501 (2006)
\bibitem{chiri} G. Chiribella, G. M. D'Ariano, P. Perinotti Phys. Rev. Lett. 101, 180504 (2008)
\bibitem{marco} Marco Lucamarini., Alessandro Cere, Giovanni Di Giuseppe, Stefano Mancini, David Vitali and Paolo Tombesi, Open Sys. \& Information Dyn. (2007) 14:169 - 178
\bibitem{lutj} N. Lutkenhaus, M. Jahma, New Journal of Physics 4 (2002) 44.1 - 44.9
\bibitem{piran} Stefano Pirandola, Stefano Mancini, Seth Lloyd and Samuel L. Braunstein, Nature Physics 4, 726 - 730 (2008) 
\bibitem{lutk} N. Lutkenhaus, Phys. Rev. A 61, 052304 (2000) 
\bibitem{phd} Xiongfeng Ma, arXiv:0808.1385v1 [quant-ph] 10 Aug 2008
\bibitem{lut99} N. Lutkenhaus, Phys. Rev. A 59, 3301 (1999) 
\bibitem{gys} C. Gobby, Z. L. Yuan, and A. J. Shields, Applied Physics Letters, Volume 84, Issue 19, pp. 3762 - 3764, (2004).
\end{thebibliography}
\end{document}